\documentclass[12pt,preprint]{aastex62}%\documentclass[12pt,preprint]{aastex63}
\begin{document}

%% You can insert a short comment on the title page using the command below.

%\slugcomment{Submitted: ApJ, January 2020}

%% If you wish, you may supply running head information, although
%% this information may be modified by the editorial offices.
%% The left head contains a list of authors,
%% usually a maximum of three (otherwise use et al.).  The right
%% head is a modified title of up to roughly 44 characters.  Running heads
%% will not print in the manuscript style.

\shorttitle{Multi-Wavelength TRGB}
\shortauthors{Madore \& Freedman}

%% This is the end of the preamble.  Indicate the beginning of the
%% paper itself with \begin{document}.

%\begin{document}


\title{\bf Mathematical Underpinnings of the
\\ Multi-Wavelength Structure of the TRGB\\}


\author{\bf Barry F. Madore} 
\affil{The Observatories \\ Carnegie
Institution for Science \\ 813 Santa Barbara St., Pasadena, CA ~~91101}
\affil{Dept. of Astronomy \& Astrophysics, University of Chicago\\ 560 S. Ellis Ave.,\\
Chicago, IL, 60637}
\email{barry.f.madore@gmail.com} 

\author{\bf Wendy L. Freedman}
\affil{Dept. of Astronomy \& Astrophysics, University of Chicago\\ 560 S. Ellis Ave.,\\
Chicago, IL, 60637}

%% are identified by the \altaffilmark after each name.  Specify  alternate

%y\medskip
%y\vfill\eject
\begin{abstract} 

We consider the application of the tip of the red giant branch
(TRGB) in the optical and in the near infrared
for the determination of distances to nearby galaxies. We analyse
ACS VI (F555W \& F814W) data and self-consistently cross-calibrate
WFC3-IR JH (F110W \& F120W) data using and absolute magnitude
calibration of $M_I = $~-4.05~mag as determined in the LMC using
detached eclipsing binary star geometric parallaxes. We demonstrate
how the optical and near-infrared calibrations of the TRGB method
are mathematically self-consistent, and illustrate the mathematical
basis and relations amongst these multi-wavelength calibrations.
We go on to present a method for determining the reddening,
extinction and the true modulus to the host galaxy using
multi-wavelength data.  The power of the method is that with
high-precision data, the reddening can be determined using the 
TRGB stars themselves, and decreases the systematic (albeit
generally small) uncertainty in distance due to reddening for these halo stars.

\end{abstract}

\keywords{ distances}
.
\medskip
\vfill\eject

\section{Introduction}

The Carnegie Chicago Hubble Program (CCHP) has been pursuing three
goals: (1) Improving the Cepheid distance scale by understanding the optical, and carefully moving to the near infrared (eg., Hoyt et al. 2018; Madore et al. 2018), and (2)
Establishing an independent and parallel path to the Hubble constant
by building a Population~II distance scale, based on RR Lyrae stars,
the tip of the red giant branch (TRGB) method (e.g., Freedman et al. 
2011; Freedman et al. 2012; Rich et al. 2014; Scowcroft et al. 
2016a,b; Beaton et al. 2016; Hatt  et al. 2017; Madore et al. 2018, 
Hatt et al. 2018a,b; Jang et al. 2018; Hoyt et al. 2018; Rich et al. 
2018; Beaton et al. 2019; Hoyt et al. 2019), and ultimately (3) An 
independent calibration of Type~Ia supernovae, leading to a
determination of $H_o$ (Freedman et al. 2019, 2020).

 The value of the Hubble constant, derived from a cosmological model (e.g., Ade et al. 2016, Planck 2018) has been at variance with
classical, locally-determined values. In
particular, studies using Cepheid variables at their base
(e.g., Freedman et al. 2001; Freedman et al. 2012; Riess et al. 2011; Riess et al. 2016; Riess et al. 2019) 
have consistently given values of the Hubble constant around 73-74~km/sec/Mpc, with 
total quoted errors being on the order of  2 to 3~km/sec/Mpc. These expansion rates are more than 3-sigma 
larger than the CMB modeling values of $H_o = 67.4 \pm 0.5$~km/sec/Mpc (Planck Collaboration 
2018). Recent results from the H0LiCOW program (Birrer, et al. 2018, Bonvin et al. 2017, 
Wong et al. 2019) using time delays from strong lensing, interpreted within the context 
of $\Lambda$CDM models, also suggest a value of $H_o$ around 
73~km/s/Mpc with errors currently running in the 2~km/s/Mpc range. A precise and accurate 
calibration of the  absolute magnitude of the TRGB as an independent means 
of establishing the astrophysical distance scale is of critical importance in resolving 
the current impasse. Interestingly, the recent TRGB calibration (Freedman et al. 2019) falls 
between the Planck and other local estimates of the Hubble constant with $H_o = 69.8 \pm 
0.8 (stat) \pm 1.7 (sys)$~km/s/Mpc.

What has not been well-recognized  is that cross all wavelengths 
the colors and absolute magnitudes of the very brightest,
first-ascent red giant-branch (RGB) stars are correlated. 
And this run of tip magnitude with intrinsic color
is well understood (and predicted) to be driven by 
well-defined and monotonic functions of 
the tip stars' atmospheric metallicity. Viewed in terms of 
their bolometric magnitudes, higher-metallicity RGB stars 
are progressively redder and brighter than low-metallicity 
stars at the same stage of their evolution (brightening at a 
rate of about 0.18 mag/dex according to Salaris \& Cassisi 
(1997). However, the wavelength-dependent, differential 
line blanketing of the cool (4,000K) stellar atmospheres of 
these luminous K giant stars, results in {\it decreasing}
luminosities with increasing colors, at wavelengths bluer 
than about 8000\AA. There is a relatively flat response of 
the tip to color in the I-band, which is followed by an 
up-turn in the trend with color at near- and mid-infrared
wavelengths.  In the color-magnitude diagram the slope 
of the red giant branch (RGB), and then the difference in 
magnitude between the horizontal branch (HB) and the giant branch 
(at some fiducial color, e.g., Sandage \& Wallerstein 1960; 
Sandage \& Smith 1966; Demarque \&  Geisler 1963; Hartwick 
1968) each have a long history of being used as metallicity
indicators of Population II systems. As mentioned above, 
this metallicity effect decreases in going from the blue to 
the visual. At red end of the optical range, in the $I$ band, 
the tip of the RGB (TRGB) has been established as one of 
the most accurate and precise methods for measuring the distances 
to nearby galaxies (e.g., Lee, Freedman \& Madore 1993; 
Jacobs et al. 2009; Hatt et al. 2017; Jang \& Lee 2017; 
Jang et al. 2018). Beyond that, at near-infrared ($JHK$) 
wavelengths where the TRGB stars are increasingly more 
luminous, a new calibration of the TRGB (Madore et al. 2018; 
Hoyt et al. 2018) shows even greater promise for extending the 
reach of the TRGB in determining distances on cosmologically
significant scales.

In this paper we make explicit the simple mathematical 
underpinning of the multi-wavelength
calibration of the TRGB. This then allows us to understand 
and apply the TRGB zero point and color-dependence calibration 
from the optical ($BVI$) and into the near-infrared (JHK), 
and then predict its behavior out into the mid-infrared (beyond 2
$\mu$m, say). Given that TRGB (spectral type K) stars are
intrinsically redder than Cepheids (having F and G spectral types), 
the absolute magnitudes of TRGB stars rapidly go from being 
fainter than Cepheids (with periods in excess of 10 days, say) at
optical ($BV$) wavelengths, to being competitive in luminosity 
in the near-infrared at 1.6 and 2.2~$\mu$m, for example. 

 We note that fitting the ``tip'' (interpreted 
to mean the brightest ``point" of the helium-flash-truncated 
red giant branch) is not a concept that applies at all wavelengths, without some modification. At all wavelengths the so-called tip 
spreads out in color as a function of metallicity, as 
it simultaneously spreads in luminosity. The TRGB rises in luminosity with increasing color in the redward bands (greater than 9000$\AA$, say), falls in the blueward bands, and holds at a fairly constant luminosity (with color/metallicity) at the transitional I-band wavelength. But the changes in luminosity for any 
given wavelength are deterministically correlated with 
the corresponding changes in color/metallicity. The near-infrared (K band) luminosity of the modern-day TRGB (as defined by the most luminous RGB stars in globular clusters) was first found, nearly 40 years ago (Frogel, Cohen \& Persson 1983), to be a brightening function of color, where they authors rightly attributed this effect to increasing metallicity (their Figures 1, 2 \& 4). More recently, with the deployment of panoramic near-infrared detectors, on the ground and in space, several groups have seen this same upward trend of NIR  luminosity with increasing color, in extragalactic populations of halo stars having a range of metallicities and colors (e.g., Dalcanton et al. 2012; Wu et al. 2014; Madore et al. 2018; Anand et al. 2019 and McQuinn et al. 2019).

%\vfill\eject

\section{Transformations}

We begin with one (linear) mathematical realization of the run of the TRGB magnitude with color:

$$M_J = a_1~(J-K)_o + b_1 ~~ . . . ~(1) $$

\par\noindent 
where $b_1$ is the $J$-band zero point,  and $a_1$ is the slope of the
TRGB in the $J$ band as a function of the intrinsic $(J-K)_o$ color. 

First, it is simple to show that the calibration of the TRGB in the $K$ band is 
pre-determined by a simple re-arrangement of terms in Equation 1,

$$M_K = M_J - (J-K)_o = a_1~(J-K)_o + b_1 - (J-K)_o
 = (a_1 - 1)~(J-K)_o + b_1$$

\par\noindent It is worth emphasizing that while the $J$ and $K$-band slopes ($a_1$
and $a_1-1$, respectively) differ by unity (exactly unity), their respective zero 
points are identical (both being equal to $b_1$); this is a mathematical tautology. 
What is less obvious is that the mapping of the calibration to other wavelengths 
is equally as simple, and just as deterministic.

For example, let us next seek out a calibration in a different bandpass, the
$H$-band, say. The key step is to establish how the new color $(J-H)_o$
quantitatively corresponds to the original $(J-K)_o$ color. Over the
small color range applicable to the TRGB it is reasonable to adopt a
linear transformation of the form

$$ (J-H)_o = \alpha (J-K)_o + \beta  ~~ . . . ~ (2)$$

\par\noindent 
Now, by definition

$$M_H = M_J - (J - H)_o$$

\par\noindent 
and then combining Equations 1 \& 2 gives

$$M_H = a_1(J-K)_o + b_1 - \alpha (J-K)_o - \beta $$

\par\noindent 
giving

$$M_H = (a_1 - \alpha)(J-K)_o + [b_1 - \beta]$$

Thus, given knowledge of one independently calibrated run of the TRGB
magnitude with color, all that is required are the color-color relations 
between different bandpasses for the small run of spectral types of stars 
populating the TRGB, in order to predict the slopes and the zero
points for all other color-magnitude combinations. 

By this same logic we then have the following calibrations for the run of the TRGB as a
function of the $(J-H)_o$ color, where $(J-K)_o = 1/\alpha (J-H)_o -\beta/\alpha, $ thus

$$M_J = [a_1/\alpha](J-H)_o + [b_1 - a_1 \beta/\alpha] $$

$$M_H = [a_1/\alpha - 1)](J-H)_o + [b_1 - a_1 \beta/\alpha] $$

$$M_K = [(a_1 - 1)/\alpha)(J-H)_o + [b_1 - (a_1 - 1)\beta/\alpha)] $$

%\vfill\eject
\section{Reddenings Self-Consistently Derived by and for the TRGB}

In one of the earliest applications of the TRGB method to extragalactic distance determinations
(e.g., Lee, Freedman \& Madore 1993) it was (plausibly) argued that reddening corrections
would be small (and/or, could be reasonably well-estimated independently) for any given line
of sight. This follows from the following two lines of argument: (1) In any given galaxy 
the old Population~II stars (of which TRGB stars are amongst the brightest components) 
make up the bulk of mass and light in the halo. Halos are generally gas- and dust-free. 
Thus by limiting observations to the halos of galaxies, any reddening corrections are 
significantly decreased as compared to the disk, for example, where Cepheids are 
located. (2) Maps of the Milky Way foreground extinction (Schlafly \& Finkbeiner 2011, 
Schlegal, Finkbeiner \& Davis 1998; Burstein \& Heiles 1982) can then be used to estimate 
the only other (independent) source of extinction affecting the observed TRGB magnitudes.

Under  extreme conditions, where the target galaxies are at low
Galactic latitude and there is high extinction (NGC~6822, IC~0342, 
Maffei~1 and 2, for example) these same maps are useful, but the
uncertainties are considerably larger. In these cases, spatial
variations may not be resolved
by the HI and IR maps that went into the tabulations, where ``sub-grid"
dust may be spatially quite patchy. Moreover, the conversion factors
(gas-to-dust ratios, etc.) that go into converting HI and dust 
re-emission may also be variable across the sky, leading to imprecise
predictions at any give position in the sky. Having a more direct means
of determining the extinction to the TRGB stars themselves would, of
course, be preferred.

To deal with these extreme cases of high extinction several novel approaches have been advanced by Tully and his collaborators: (1) For galaxies close to the Galactic plane the statistical color properties of the abundance of foreground red giant branch stars provide an estimation of line-of-sight reddening, as introduced and applied in Anand et al. (2019). (2) For highly-extincted galaxies, that also happen to have on-going star formation in their disks, then the redward displacement of the zero age main sequence can be used to estimate the total line-of-sight extinction. This provides an upper limit to the TRGB reddening given that the disk stars will also be suffering from {\it in situ} dust extinction that will not apply to the TRGB stars in the halo of that same galaxy.

Finally, for cases of intermediate extinction, such as the LMC, where comparable amounts of internal and foreground extinction are arising for a more centrally located (bar) population of TRGB stars, a differential, multi-wavelength comparison of observed tip magnitudes in the target galaxy (the LMC) with the corresponding tip magnitudes in a very low-reddening comparison galaxy (the SMC or IC~1613, for example) provide another path. In this application, high-precision tip detections, are required in at least three well-separated wavelength bands, resulting in the run of apparent distance moduli as a function of wavelength. These can then be fit by an adopted extinction curve, where the only remaining free parameter is the total line-of-sight extinction. This method is a generalization of a technique regularly used to de-redden extragalactic Cepheids (for a recent application, in that context, see Rich et al. 2018), and it has been successfully applied to the TRGB zero-point calibration recently published for the LMC by Freedman et al. (2020).

Ultimately, having a variety of estimates of the total line-of-sight reddening will, at very least, provide a better estimate of the systematic uncertainty to be associated with this correction. And, as it has been stated many times before, in many similar contexts, by moving as far into the infrared as possible, without sacrificing precision, will always reduce the impact of extinction, which continuously falls in its magnitude with increasing wavelength.

However, before proceeding further we provide visualizations of the coupled equations describing the 
interdependence of the slopes, zero points and mean magnitudes of the TRGB calibration 
across bands. Figures 1 \& 2 show the calibrating equations for the near-infrared against 
a backdrop of $JHK$ color-magnitude data for the galaxy IC~1613. The IC~1613 data are 
presented in Madore et al. (2018). Figure 3 is a simplified representation of the 
interconnected calibrations, going from the downward slanting $V$-band calibration, 
through the nearly-flat $I$-band TRGB, and then out into the near-infrared, where all 
of the calibrations trend upward with increasing color. This figure simultaneously illustrates 
the monotonic increase of the mean magnitude level of the TRGB population as a 
function of increasing wavelength. These multi-wavelength trends and correlations 
were all first illustrated in Figures 12 and 13 of Madore et al. (2018).

With multi-wavelength observations and their corresponding intrinsic calibrations of the
TRGB, it is now possible to consider an alternative means of determining extinction
corrections for the TRGB method using the tip stars themselves and for the same
lines of sight being used for each galaxy individually. We now offer a first
look at that possibility below.

\subsection {The Formulae for Determining Reddenings}

If we have three band passes ($JHK$, say) from which we can determine two independent
{\it apparent} TRGB distance moduli, then it is possible to extract the reddening from the
difference in these two moduli. However, because of the sloping nature of the
color-magnitude description of the trace of the TRGB, the difference in apparent moduli is
not simply the color excess. Using Figure 3 as a guide we can understand that the 
apparent distance modulus will be composed of {\it three terms}: $\mu_o$, the true distance 
modulus, $A_{\lambda}$, the extinction at the measured wavelength, and an additional 
term that numerically is the reddening multiplied by the slope $a_{\lambda}$ of the 
intrinsic TRGB color-magnitude calibration. For instance

  $$\mu_J = b_J - b_1 = \mu^o_J + A_J + a_1 E(J - H) $$
 
 and
 
  $$\mu_K = b_K - (b_1 - \beta) = \mu^o_K + A_K + (a_1 - \alpha) E(J - H) $$

Recognizing that $\mu^o_J = \mu^o_K$, this gives 

  $$ \Delta\mu(JK) = \mu_J - \mu_K = b_J - b_K + \beta $$
  $$ = A_J + a_1 E(J - H) - A_K - (a_1 - \alpha) E(J - H)$$
  $$ = E(J - K) + \alpha E(J - H)$$
  $$ = X_{JHK} E(J - H) + \alpha E(J - H) = (\alpha + X_{JHK}) E(J - H)$$
where 
  $$ X_{JHK}  = E(J - K)/E(J - H) $$
  
  giving
  
  $$E(J - H) = \Delta\mu(JK)/(\alpha + X_{JHK})$$

This means that, for the first time, it is in principle possible to determine
directly and self-consistently the total line-of-sight extinction to the
TRGB field using the TRGB stars themselves.

In the event that even more bandpasses are introduced, the value of the
reddening will become over-constrained. At this point a
scaled-extinction-curve fit to the apparent moduli as a function of
wavelength will give a mean reddening and true modulus, as has been
frequently done for determining these same parameters using
multi-wavelength observations of extragalactic Cepheids (as first done by Freedman  
1988; and more recently in  Rich et al. 2014). Indeed, examples of an 
application of this approach using the near-infrared TRGB method can be seen in the 
simultaneous determination of the distance and reddening to the LMC, SMC and IC~1613 in Freedman et al. (2020).

\begin{figure*} \centering \includegraphics[width=18.0cm, angle=-0]
{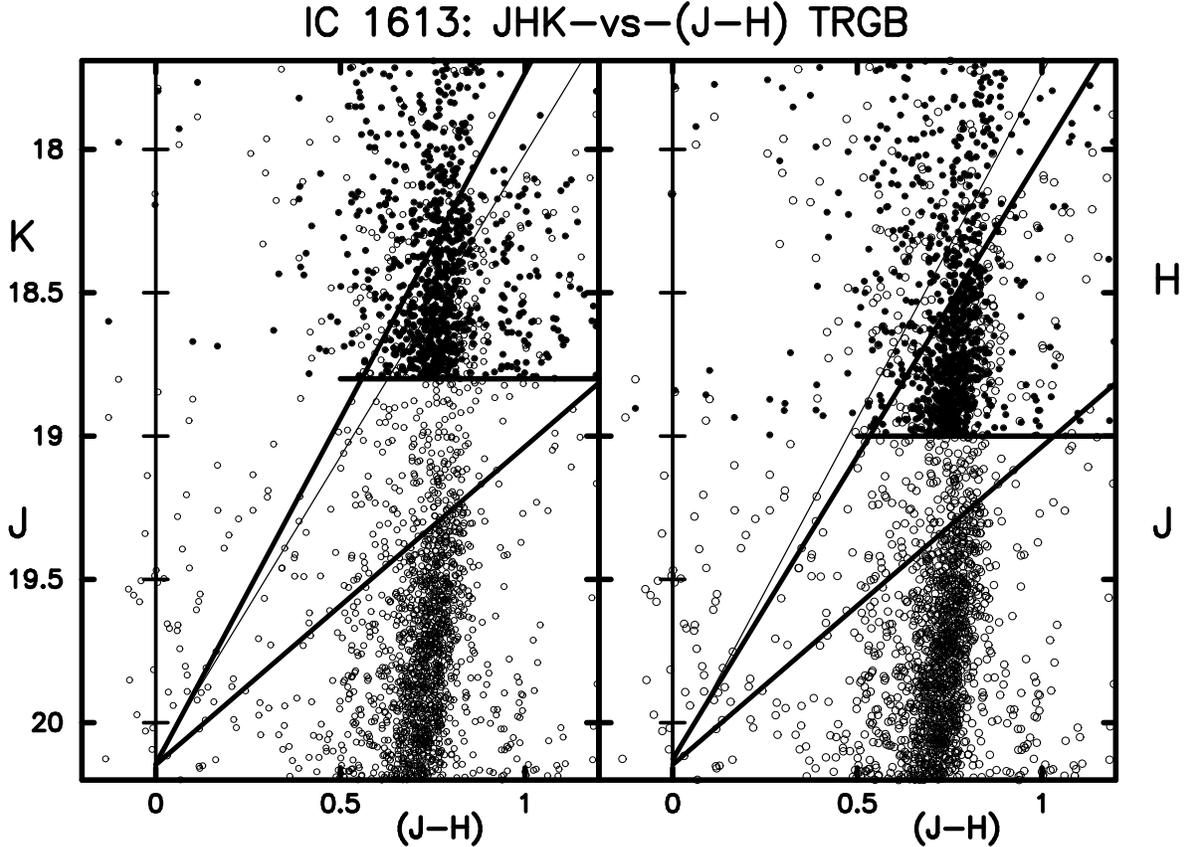} \caption{\small A montage of $J,H,K$
vs $(J-H)$ color-magnitude diagrams for IC~1613, centered on the upper
magnitude of the TRGB and red giant branch. In the upper portion of the
left plot is the $K$-band TRGB/CMD (small solid dots) truncated at $K = $
18.8~mag. Upon this is superimposed the full $J$-band CMD (small open
circles). The thick solid black lines mark the sloping TRGB,
extrapolated to $(J-H) =$ 0.00~mag. The steeper line is a fit to the
$K$-band data; the shallower line is a fit to the $J$-band TRGB. For clarity
the $H$-band data and fits are presented in the right panel where the
$J$-band data are reproduced.  Thin solid lines give the fits to the
data in alternate panels [i.e., the thin line in the left (K-band) panel is a muted rendering of the thick line (H-band) solution seen in the right panel, and vice versa] emphasizing how all three lines converge to the same Y intercept at (J-H) = 0.0~mag}
\end{figure*}

\begin{figure*} \centering \includegraphics[width=18.0cm, angle=-0]
{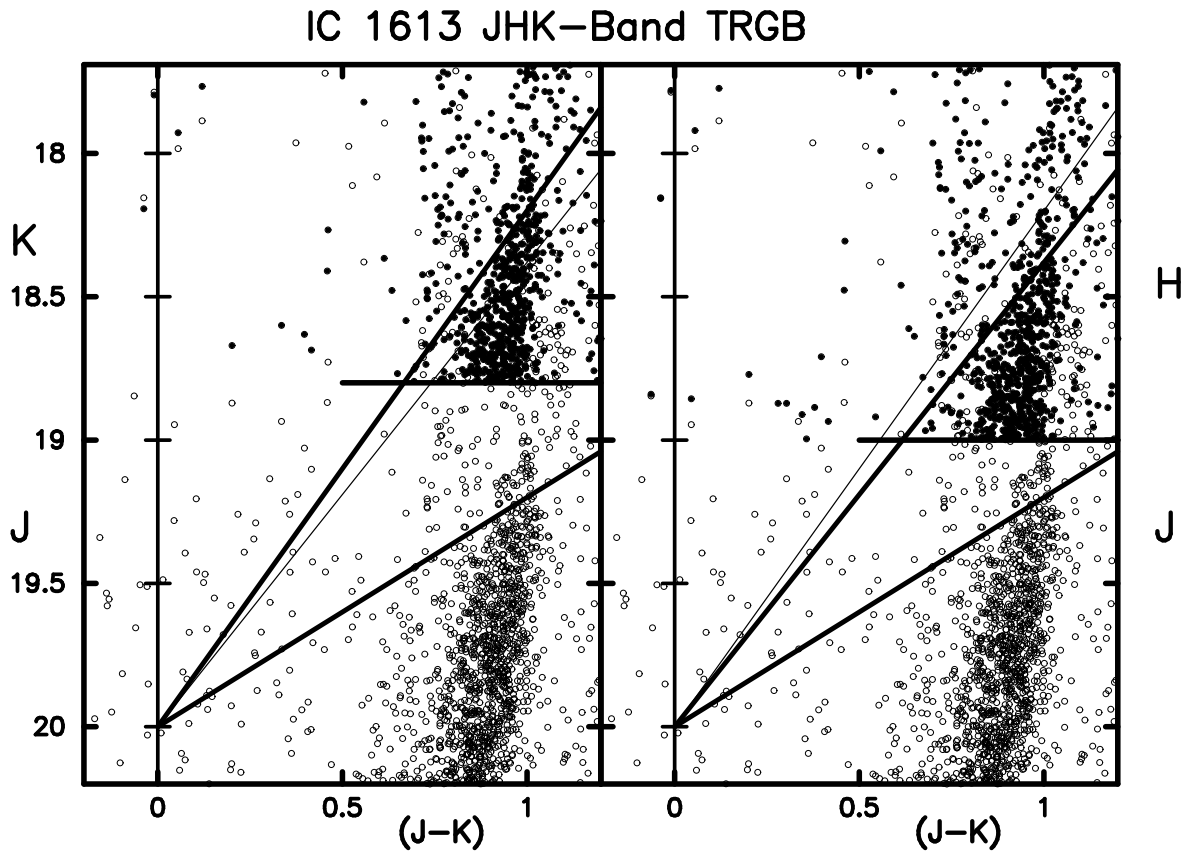} \caption{\small The same as Figure 1
in all respects except for a change in the color index. In this case the
(J-K) color is being used. } \end{figure*}

\begin{figure*} \centering \includegraphics[width=12.0cm, angle=-0]
{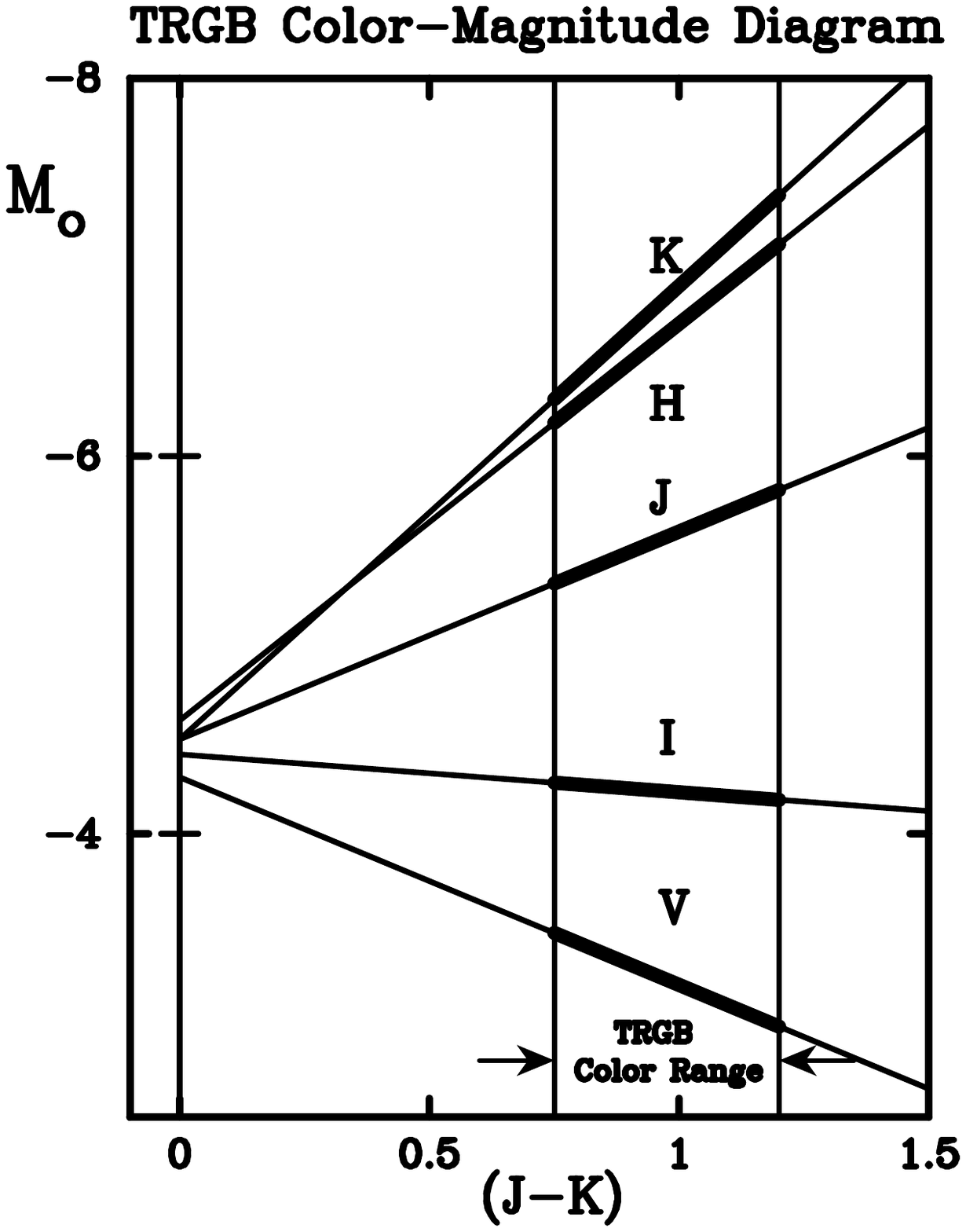} \caption{\small A purely graphical
representation of the coordinated brightening of the absolute magnitude in step with 
the increasing slope of the TRGB with the increasing wavelength of the absolute
magnitude being considered. The slopes and zero points are those of the
calibrations presented in Madore et al. (2018) for the near-infrared
calibration of the TRGB for IC~1613. See also Figure 1. } \end{figure*}

\vfill\eject
\section{Summary and Conclusions}

We have explicitly provided the mathematical underpinnings of the color
dependence of the magnitude of the TRGB as a function of wavelength and metallicity. 
In doing so we have emphasized the fact that for a given color the slopes and 
zero-point magnitudes are predetermined by the (distance-independent) intrinsic color-color 
relations between the various band passes being considered. It is 
a generic property of the TRGB calibration that at progressively longer wavelengths,
the absolute magnitudes of the stars defining the TRGB will become brighter, 
in lock step with the slopes of the color-magnitude relation becoming steeper: 
longer wavelengths, steeper slopes, brighter absolute magnitudes. This simple 
property then provides a means to self-consistently correct TRGB magnitudes 
and distances for total line-of-sight reddening and extinction.

%\vfill\eject
\section{Acknowledgements} We thank the {\it Observatories of the Carnegie Institution for
Science} and the {\it University of Chicago} for their support of our long-term research
into the calibration and determination of the expansion rate of the Universe. The
near-infrared observations discussed here were taken with the FourStar camera on the Baade
6.5m telescope at the Las Campanas Observatory, Chile. Support for this work was provided
in part by NASA through grant number HST-GO-13691.003-A from the Space Telescope Science
Institute, which is operated by AURA, Inc., under NASA contract NAS 5-26555.  We thank Kayla Owens for her comments on a final draft of this paper; and finally, we thank the referee, Brent Tully, for his insightful perspectives on this method and its application.

\vfill\eject
\section{References}
\medskip

\noindent
 Anand, A.S., Tully, R.B., Rizzi, L. et al. 2019, ApJ, 880, 52

\noindent
Beaton, R.L., Freedman, W.L., Madore, B.F. et al. 2016, ApJ., 832, 210

\noindent
Beaton, R.L., Seibert, M., Hatt, D. et al. 2019, ApJ, 885, 141

\noindent
Birrer, S., Treu, T., Rusu, C.E. et al. 2018, MNRAS, 484, 4726

\noindent
Bonvin, V., Courbin, F., Suyu, S.H. et al. 2017, MNRAS, 465, 4914

\noindent
Burstein, D. \& Heiles, C. 1982, AJ, 87, 1165

\noindent
Dalcanton, J.J., Williams, B.F., Seth, A.C. et al. 2009, ApJS, 183, 67

\noindent
 Dalcanton, J.J., Williams, B.J., Melbourne, J.L. et al. 2012, ApJS, 198, 6

\noindent
Demarque, P. \& Geisler, J.E. 1963, ApJ, 137, 1102

\noindent
Freedman, W.L. 1988, ApJ, 326, 691 

\noindent
Freedman, W.L., Madore, B.F., Gibson, B.K. et al. 2001, ApJ, 553, 47

\noindent
Freedman, W.L., Madore, W.L., Scowcroft, V. et al. 2011, AJ, 142, 194

\noindent
Freedman, W.L., Madore, B.F., Scowcroft, V. et al. 2012, ApJ, 758, 24

\noindent
Freedman, W.L., Madore, B.F., Hatt, D., 2019, ApJ, 882, 34

\noindent
Freedman, W.L., Madore, B.F., Hoyt, D., 2020, ApJ, 891, 57 arXiv: 2002.01550

\noindent
Frogel, J., Cohen, J., \& Persson, S.E. 1983, ApJ, 275, 773

\noindent
Hartwick, F.D.A. 1968, ApJ, 154, 475

\noindent
Hatt, D., Beaton, R.L.,Freedman, W.L.,  et al. 2017, ApJ, 845, 146

\noindent
Hatt, D., Freedman, W.L., Madore, B.F., et al. 2018a, ApJ, 861, 104

\noindent
Hatt, D., Freedman, W.L., Madore, B.F., et al. 2018b, ApJ, 866, 145

\noindent
Hoyt, T., Freedman, W.L., Madore, B.F.  et al. 2018, ApJ, 858, 12

\noindent
Hoyt, T., Freedman, W.L., Madore, B.F.  et al. 2019, pJ, 882, 150

\noindent
Jacobs, B.A., Rizzi, L., Tully, R,B., et al. 2009, AJ, 138, 322 

\noindent
Jang, I.-S., Hatt, D., Beaton, R.L. et al. 2018, ApJ, 852, 60

\noindent
Jang, I.-S. \& Lee, M.-G. 2017, ApJ, 836, 74

\noindent
Lee, M.-G., Freedman, W.L., \& Madore, B.F. 1993, ApJ, 417, 553 

\noindent
Madore, B.F., Freedman, W.L., Hatt, D. et al. 2018, ApJ, 858, 11

\noindent
McQuinn, K.B.W., Boyer, M., Skillman, E.D. et al. 2019, ApJ, 880, 63

\noindent
Planck Collaboration, Aghanim, N., Akrami, Y., et al. 2018,
arXiv:1807.06209

\noindent
Rich, J., Persson, S.E., Freedman, W.L., Madore, B.F. et al. 2014, ApJ, 794, 107

\noindent
Rich, J.A., Madore, B.F., Monson, A.J. et al. 2018, ApJ, 869, 82  

\noindent
Riess, A.G., Macri, L.,  Cassertano, S.  et al. 2011, ApJ, 730, 119

\noindent
Riess, A.G., Macri, L., Hoffmann, S.L. et al. 2016, ApJ, 826, 56 

\noindent
Riess, A.G., Cassertano, S., Yuan, W.  et al. 2019, ApJ, 875, 145

\noindent
Salaris, M. \& Cassisi, S. 1997, MNRAS, 289, 406, 

\noindent
Sandage, A.R.  \& Wallerstein, G. 1960, ApJ, 131, 598

\noindent
Sandage, A.R. \& Smith, L.L. 1966, ApJ, 144 , 886

\noindent
Schlafly, E.F. \& Finkbeiner, D.P. 2011, ApJ, 737, 103

\noindent
Schlegel, D.J., Finkbeiner, D.P. \& Davis, M. 1998, ApJ, 500, 525

\noindent
Scowcroft, V., Freedman, W.L., Madore, B.F. et al. 2016a, ApJ, 816, 49 

\noindent
Scowcroft, V., Seibert, M., Freedman, W.L. et al. 2016b, MNRAS, 459, 1170 

\noindent
Wu, P.-F., Tully, R.B., Rizzi, L., et al. 2014, AJ, 148, 7

\noindent
Wong, K. C., Suyu, S.,  H., Chen, G.C., et al. 2020, MNRAS, XXX, 1661, arXiv:1907.04869

\vfill\eject
\end{document}